\documentstyle[preprint,aps]{revtex}
\widetext
\draft
\tighten

\begin{document}

\title{Double (implicit and explicit) dependence of the electromagnetic
field of an accelerated charge on time: Mathematical and physical analysis
of the problem}

\bigskip

\author{\bf Andrew E. Chubykalo and Stoyan J. Vlaev}

\address {Escuela de F\'{\i}sica, Universidad Aut\'onoma de Zacatecas \\
Apartado Postal C-580\, Zacatecas 98068, ZAC., M\'exico}

\date{\today}

\maketitle


\baselineskip 7mm

\begin{abstract}
We considered the electromagnetic field of a charge moving with
a constant acceleration along an axis. We found that this field
obtained from the Li\'enard-Wiechert potentials does not satisfy Maxwell
equations if one considers exclusively a retarded
interaction (i.e. pure implicit dependence this field on time). We show
that if and only if one takes into account both retarded interaction and
direct interaction (so called ``action-at-a-distance") the field produced
by an accelerated charge satisfies Maxwell equations.
\end{abstract}


$$$$
{\bf 1. Introduction}

The problem of a calculation of the potentials and the fields  created by a
point charge moving with an acceleration  was raised
for the first time about 100 years ago by A.Li\'enard and E.Wiechert [1]
and has not lost relevance nowadays.  The question concerning the
choice of a correct way to obtain these fields - seems to have been solved
finally (see, e.g.  well-known book by L.D.Landau [2]). But lately many
authors (see e.g. [3-6] and others in References,  and this list one could
continue) have time and again resorted to this problem - a problem which
was given up by contemporary physics long ago. We think that there must
be something behind it that makes the problem still actual from both the
scientific  and  pedagogical points of view.

It is well-known that the electromagnetic field created by an arbitrarily
moving charge
   \begin{equation}
   {\bf E}({\bf r},t)=q\left\{\frac{({\bf R}-R\frac{{\bf
         V}}{c})(1-\frac{V^{2}}{c^{2}})}{(R-{\bf R}\frac{{\bf
   V}}{c})^{3}}\right\}_{t_0}+q\left\{\frac{[{\bf R}\times[({\bf
   R}-R\frac{{\bf V}}{c})\times\frac{{\bf{\dot{V}}}}{c^{2}}]]}{(R-{\bf
   R}\frac{{\bf V}}{c})^{3}}\right\}_{t_0},
   \end{equation}
\begin{equation}
{\bf B}({\bf r},t)=\left\{\left[\frac{{\bf R}}{R}\times{\bf
E}\right]\right\}_{t_0}
\end{equation}

was obtained directly from Li\'enard-Wiechert
potentials [2]:

\begin{equation}
\varphi({\bf r},t)=\left\{\frac{q}{\left(R-{\bf R}\frac{{\bf
   V}}{c}\right)}\right\}_{t_0},\qquad
{\bf A}({\bf r},t)=\left\{\frac{q{\bf V}}{c\left(R-{\bf R}\frac{{\bf
   V}}{c}\right)}\right\}_{t_0}.
\end{equation}
The notation $\bigl\{...\bigr\}_{t_0}$ means that all functions of
$x,y,z,t$ into the parenthesis \{\} are taken at the moment of time
$t_0(x,y,z,t)$ [2] (the instant
 $t_0$ is determined from the condition (8), see below).

  Usually, the first terms of the right-hand sides ({\it rhs}) of (1) and
 (2) are called ``velocity fields" and the second ones are called
 ``acceleration fields".

It was recently
claimed by E.Comay [7] that ``...  {\it Acceleration
fields by themselves do not satisfy Maxwell's equations} [8]. {\it Only
the sum of velocity fields and acceleration fields satisfies Maxwell's
equations}." We wish to argue that this sum {\it does not satisfy}
Maxwell's equations
\begin{eqnarray} && \nabla\cdot{\bf
E}=4\pi\varrho,\\ && \nabla\cdot{\bf B}=0,\\ && \nabla\times{\bf
B}=\frac{4\pi}{c}{\bf j}+\frac{1}{c}\frac{\partial {\bf E}}{\partial t},\\
&& \nabla\times{\bf E}=-\frac{1}{c}\frac{\partial {\bf B}}{\partial t}
\end{eqnarray}
in the case when one takes into consideration  {\it exclusively} a
retarded interaction.

The remainder of our paper is organized as follows:
In Section 2 we derive the fields {\bf E} and {\bf B} taking into account
exclusively the {\it implicit} dependence the potentials $\varphi$ and
{\bf A} on time $t$. In Section 3 we prove that the field
obtained from the Li\'enard-Wiechert potentials does not satisfy Maxwell
equations if one considers exclusively a retarded
interaction (in the other words, the {\it implicit} dependence the
potentials on time of observation $t$ only). In Section 4 we consider
another way to obtain the fields {\bf E} and {\bf B}. This way is
based on a different type of calculation of the derivatives
$\partial\{\}/\partial t$ and $\partial\{\}/\partial x_i$ in which
the functions $\varphi$ and ${\bf A}$ are considered as
functions with a {\it double} dependence on $(t,x,y,z)$: implicit and
explicit {\it simultaneously}. By this way one  obtains formally the {\it
same} expressions (1) and (2) for the fields.  If one uses {\it
this} manner to verify the validity of Maxwell's equations, one finds that
fields (1) and (2) satisfy these equations. In this Section  we
shall show that this way does  not correspond to a {\it pure} retarded
interaction between the charge and the point of observation. Section 5
closes the paper.

$$$$
{\bf 2. Deriving the fields E and B taking into account
the retarded interaction only}

Let us try to derive the formulas (1), (2) for
the electric ({\bf E}) and magnetic ({\bf B}) fields  taking into
account that the {\it state} of the fields {\bf E} and {\bf B} at the
instant $t$ must be {\it completely} determined by the {\it state} of the
charge at the instant $t_0$. The instant
 $t_0$ is determined from the condition (see Eq.(63.1) of Ref.[2]):
 \begin{equation}
 t_0=t-\tau=t-\frac{R(t_0)}{c}.
 \end{equation}
Here $\tau=R(t_0)/c$ is the so
 called ``retarded time", $R=|{\bf R}|$, {\bf R} is the vector  connecting
the site ${\bf r}_0(x_0,y_0,z_0)$ of the charge $q$ at the instant $t_0$
with the point of observation {\bf r}($x,y,z$).

 All the quantities on the {\it rhs} of (3) must be evaluated at the time
 $t_0$ (see [2], the text after Eq.(63.5)), which, in turn, depends on
 $x,y,z,t$:

 \begin{equation}
t_0=f(x,y,z,t).
 \end{equation}

Let us, to be more specific, turn to Landau and Lifshitz who write ([2],
p.161)\footnote{We use here our numeration of formulas: our (3) is (63.5)
of [2], (8) is (63.1) of [2].}:

``{\it To calculate the intensities of the electric and magnetic fields
from the formulas
\begin{equation} {\bf
E}=-\nabla\varphi-\frac{1}{c}\frac{\partial{\bf A}}{\partial t},\qquad
{\bf B}=[\nabla\times{\bf A}].
\end{equation}
we must differentiate
$\varphi$ and {\bf A} with respect to the coordinates x, y, z of the
point, and the time t of observation. But the formulas (3) express the
potentials as a functions of $t_0$}, {\bf and only through} {\it the
relation (8)} {\bf as implicit} {\it functions of x, y, z, t. Therefore
to calculate the required derivatives we must first calculate the
derivatives of $t_0$}".

Now, following this note of Landau, we can construct a scheme of
calculating the required derivatives, taking into account that $\varphi$
and {\bf A} {\it must not} depend on $x,y,z,t$ {\it explicitly}:
\begin{equation}
\left.
{\Large
\begin{array}{c}
{
\frac{\partial\varphi}{\partial x_i}=
\frac{\partial\varphi}{\partial t_0}\frac{\partial t_0}{\partial
x_i} }\\ { }\\ {\frac{\partial{\bf A}}{\partial t}= \frac{\partial{\bf
A}}{\partial t_0}\frac{\partial t_0}{\partial t} }\\ { }\\
{\frac{\partial A_k}{\partial x_i}=
\frac{\partial A_k}{\partial t_0}\frac{\partial t_0}{\partial x_i}
}
\end{array}}
\right\}
\end{equation}

To obtain Eqs. (1) and (2), let us rewrite Eqs.(10) taking into account
Eqs.(11)\footnote{In Eqs. (12),(13)
we have used the well-known formulas of the vectorial analysis:  $$
\nabla u= \frac{\partial u}{\partial\xi}\nabla\xi \qquad {\rm and}\qquad
[\nabla\times{\bf f}]=\left[\nabla\xi\times\frac{\partial{\bf f}}{\partial
\xi}\right] $$
where $u=u(\xi)$,
${\bf f}={\bf f}(\xi)$ and $\xi=\xi(x,y,z)$.}:

\begin{equation}
{\bf E}=-\nabla\varphi-\frac{1}{c}\frac{\partial{\bf A}}{\partial
t}=-\frac{\partial\varphi}{\partial t_0}\nabla
t_0-\frac{1}{c}\frac{\partial {\bf A}}{\partial t_0}\frac{\partial
t_0}{\partial t},
\end{equation}
\begin{equation}
{\bf B}=[\nabla\times{\bf A}]=\left[\nabla t_0\times\frac{\partial {\bf
A}}{\partial t_0}\right].
\end{equation}

To calculate Eqs.(12),(13) we use relations
$\partial t_0/\partial t$ and $\partial t_0/\partial x_i$ obtained in [2]:
\begin{equation}
\frac{\partial t_0}{\partial t}=\frac{R}{R-{\bf RV}/c}
\qquad
{\rm and}\qquad
\frac{\partial t_0}{\partial x_i}=-
\frac{x_i-x_{0i}}{c[R-{\bf RV}/c]}.
\end{equation}

From Eqs.(3) we find:
\begin{equation}
\frac{\partial\varphi}{\partial t_0}=-\frac{q}{(R-{\bf R}
\mbox{\boldmath$\beta$})^2}\left(\frac{\partial R}{\partial
t_0}-\frac{\partial{\bf R}}{\partial t_0}\mbox{\boldmath$\beta$}-{\bf
R}\frac{\partial\mbox{\boldmath$\beta$}}{\partial t_0}\right),
\end{equation}
where $\mbox{\boldmath$\beta$}={\bf V}/c$. Hence, taking into account
that\footnote{This follows from the expressions $R=c(t-t_0)$ and ${\bf
R}={\bf r}-{\bf r}_0(t_0)$. See [2].} $$ \frac{\partial
R}{\partial t_0}=-c,\qquad\frac{\partial{\bf R}}{\partial
t_0}=-\frac{\partial{\bf r}_0}{\partial t_0}=-{\bf V}(t_0)\qquad {\rm and}
\qquad\frac{\partial{\bf V}}{\partial t_0}=\dot{\bf V},
$$
we have (after an algebraic simplification):
\begin{equation}
\frac{\partial\varphi}{\partial t_0}=\frac{qc(1-\beta^2+ {\bf
R}\dot{\mbox{\boldmath$\beta$}}/c)}{(R-{\bf R}
\mbox{\boldmath$\beta$})^2}.
\end{equation}
In  turn
\begin{equation}
\frac{\partial {\bf A}}{\partial t_0}=\frac{\partial\varphi}{\partial t_0}
\mbox{\boldmath$\beta$}+\varphi\dot{\mbox{\boldmath$\beta$}}.
\end{equation}
Putting $\varphi$ from Eqs.(3), Eq.(16) and Eq.(17) together, we have
(after simplification):

\begin{equation}
\frac{\partial {\bf A}}{\partial
t_0}=qc\,\,\frac{\mbox{\boldmath$\beta$}(1- \beta^2+ {\bf
R}\dot{\mbox{\boldmath$\beta$}}/c)+
(\dot{\mbox{\boldmath$\beta$}}/c)(R-{\bf R}
\mbox{\boldmath$\beta$})}{(R-{\bf R}
\mbox{\boldmath$\beta$})^2}.
\end{equation}
Finally, substituting Eqs. (14), (16) and (18) in Eq.(12) we obtain:
\begin{eqnarray}
{\bf E} &=& \frac{qc(1-\beta^2+ {\bf
R}\dot{\mbox{\boldmath$\beta$}}/c)}{(R-{\bf R}
\mbox{\boldmath$\beta$})^2}\left(-\frac{{\bf R}}{c(R-{\bf
R}\mbox{\boldmath$\beta$})}\right)- \nonumber\\
&& \nonumber\\
&& -\,\,q\,\,\frac{\mbox{\boldmath$\beta$}(1- \beta^2+ {\bf
R}\dot{\mbox{\boldmath$\beta$}}/c)+
(\dot{\mbox{\boldmath$\beta$}}/c)(R-{\bf R}
\mbox{\boldmath$\beta$})}{(R-{\bf R}
\mbox{\boldmath$\beta$})^2}\left(\frac{R}{R-{\bf
R}\mbox{\boldmath$\beta$}}\right)= \nonumber\\
&& \nonumber\\
&& = q\,\,\frac{{\bf R}(1-\beta^2+ {\bf
R}\dot{\mbox{\boldmath$\beta$}}/c)-R\mbox{\boldmath$\beta$} (1-\beta^2+
{\bf
R}\dot{\mbox{\boldmath$\beta$}}/c)-(R\dot{\mbox{\boldmath$\beta$}}/c)(R-
{\bf R}\mbox{\boldmath$\beta$})}{(R-{\bf R}\mbox{\boldmath$\beta$})^3}.
\end{eqnarray}

Grouping together all terms with acceleration, one can reduce
this expression to \begin{equation} {\bf E}=q\,\frac{({\bf R}-R\frac{{\bf
         V}}{c})(1-\frac{V^{2}}{c^{2}})}{(R-{\bf R}\frac{{\bf
   V}}{c})^{3}}+q\,\frac{({\bf
R}\dot{\mbox{\boldmath$\beta$}}/c)({\bf R}-R\mbox{\boldmath$\beta$})-
(R\dot{\mbox{\boldmath$\beta$}}/c)(R- {\bf R}\mbox{\boldmath$\beta$})}{
(R-{\bf R}\mbox{\boldmath$\beta$})^3}.
\end{equation}
Now, using the formula of double vectorial product, it is not worth
reducing the numerator of the second term of Eq.(20) to  $[{\bf
R}\times[({\bf R}-R\mbox{\boldmath$\beta$})\times
\dot{\mbox{\boldmath$\beta$}}/c]]$. As a result we have Eq.(1).

Analogically, substituting Eqs. (14) and (18) in Eq.(13) we obtain
\begin{equation}
{\bf B}=\left[\frac{{\bf R}}{R}\times q\,\frac{
-R\mbox{\boldmath$\beta$} (1-\beta^2+
{\bf
R}\dot{\mbox{\boldmath$\beta$}}/c)-(R\dot{\mbox{\boldmath$\beta$}}/c)(R-
{\bf R}\mbox{\boldmath$\beta$})}{(R-{\bf R}\mbox{\boldmath$\beta$})^3}\right].
\end{equation}
If we add  ${\bf R}(1-\beta^2+ {\bf
R}\dot{\mbox{\boldmath$\beta$}}/c)$ to the numerator of the second term of
the vectorial product (21)\footnote{The meaning of Eq.(21) does not change
because of $[{\bf R}\times{\bf R}]=0$.} we obtain Eq.(2) (see
Eq.(19))

In the next section we shall consider a charge moving with a constant
acceleration along the $X$ axis and we shall show that the Eq.(7) {\it is
not satisfied} if one substitutes ${\bf E}$ and ${\bf B}$ from Eqs.(1) and
(2) in Eq.(7) and {\it takes into consideration exclusively a retarded
interaction}.  To verify this we have to find the derivatives of
$x-,y-,z-$components of the fields ${\bf E}$ and ${\bf B}$ with respect to
the time $t$ and the coordinates $x,y,z$.  The functions ${\bf E}$ and
${\bf B}$ depend on $x,y,z,t$ through $t_0$ from the conditions (8)-(9).
In other words, we shall show that {\it these fields ${\bf E}$ and ${\bf
B}$ do not satisfy the Maxwell equations if the differentiation rules (11)
that were applied to $\varphi$ and ${\bf A}$ (to obtain ${\bf E}$ and
${\bf B}$) are applied identically to ${\bf E}$ and ${\bf B}$}.

$$$$
{\bf 3. Does the} {\it retarded} {\bf electromagnetic field of a charge
moving with a constant acceleration satisfy Maxwell equations?}

Let us consider a charge $q$ moving with a constant acceleration
along the $X$ axis. In this case its velocity and acceleration have only
$x$-components, respectively ${\bf V}(V,0,0)$ and ${\bf a}(a,0,0)$.
Now we  rewrite the Eqs. (1) and (2) by components:

\begin{equation}
E_x(x,y,z,t)=q\left\{\frac{(V^2-c^2)[RV-c(x-x_0)]}{\left[(cR-
V(x-x_0)\right]^3}\right\}_{t_0}   + q\left\{
\frac{ac[(x-x_0)^2-R^2]}{\left[(cR-V(x-x_0)\right]^3}\right\}_{t_0},
\end{equation}

\begin{equation}
E_y(x,y,z,t)=-q\left\{\frac{c(V^2-c^2)(y-y_0)}{\left[(cR-V(x-
x_0)\right]^3}\right\}_{t_0} + q\left\{
\frac{ac(x-x_0)(y-y_0)}{\left[(cR-V(x-x_0)\right]^3}\right\}_{t_0},
\end{equation}

\begin{equation}
E_z(x,y,z,t)=-q\left\{\frac{c(V^2-c^2)(z-z_0)}{\left[(cR-V(x-
x_0)\right]^3}\right\}_{t_0} + q\left\{
\frac{ac(x-x_0)(z-z_0)}{\left[(cR-V(x-x_0)\right]^3}\right\}_{t_0},
\end{equation}

\begin{equation}
B_x(x,y,z,t)=0,
\end{equation}

\begin{equation}
B_y(x,y,z,t)=q\left\{\frac{V(V^2-c^2)(z-z_0)}{\left[(cR-V(x-
x_0)\right]^3}\right\}_{t_0} - q\left\{
\frac{acR(z-z_0)}{\left[(cR-V(x-x_0)\right]^3}\right\}_{t_0},
\end{equation}

\begin{equation}
B_z(x,y,z,t)=-q\left\{\frac{V(V^2-c^2)(y-y_0)}{\left[(cR-
V(x-x_0)\right]^3}\right\}_{t_0} + q\left\{
\frac{acR(y-y_0)}{\left[(cR-V(x-x_0)\right]^3}\right\}_{t_0},
\end{equation}

\medskip

Obviously, these components are functions of $x,y,z,t$  through $t_0$ from
the conditions (8)-(9). This means that when we substitute the field
components given by Eqs.(22)-(27) in the Maxwell equations (4)-(7), we
once again have to use the differentiation rules as in (11):

\medskip

\begin{equation}
\left.
{\Large
\begin{array}{c}
{\frac{\partial E\{{\rm or}\,B\}_k}{\partial t}=
\frac{\partial E\{{\rm or}\, B\}_k}{\partial t_0}\frac{\partial
t_0}{\partial t}, }\\
{
 } \\
{\frac{\partial E\{{\rm or}\,B\}_k}{\partial x_i}=
\frac{\partial E\{{\rm or}\, B\}_k}{\partial t_0}\frac{\partial
t_0}{\partial x_i},
}
\end{array}}
\right\}
\end{equation}
\medskip
where $k$ and $x_i$ are $x,y,z$.

Remember that we are considering the case with ${\bf V}=(V,0,0)$, so,
 one obtains:
\begin{equation}
\frac{\partial t_0}{\partial t}=\frac{R}{R-(x-x_0)V/c}
\qquad
{\rm and}\qquad
\frac{\partial t_0}{\partial x_i}=-
\frac{x_i-x_{0i}}{c[R-(x-x_0)V/c]}.
\end{equation}.

Let us rewrite Eq.(7) by components taking into account the rules (28) and
Eq.(25):

\medskip

\begin{equation}
\frac{\partial E_z}{\partial t_0}\frac{\partial t_0}{\partial y}
-\frac{\partial
E_y}{\partial t_0}\frac{\partial t_0}{\partial z}=0,
\end{equation}

\begin{equation}
\frac{\partial E_x}{\partial t_0}\frac{\partial t_0}{\partial z}
-\frac{\partial
E_z}{\partial t_0}\frac{\partial t_0}{\partial x}+
\frac{1}{c}\frac{\partial B_y}{\partial t_0}\frac{\partial t_0}
{\partial t}
=0,
\end{equation}

\begin{equation}
\frac{\partial E_y}{\partial t_0}\frac{\partial t_0}{\partial x}
-\frac{\partial
E_x}{\partial t_0}\frac{\partial t_0}{\partial y}+
\frac{1}{c}\frac{\partial B_z}{\partial t_0}\frac{\partial t_0}
{\partial t}
=0.
\end{equation}

\medskip

In order to calculate the derivatives $\partial E({\rm or}\,B)_k/\partial
t_0$ we need the values of the expressions  $\partial V/\partial
t_0$, $\partial x_0/\partial t_0$ and $\partial R/\partial t_0$.  In our
case we have to use\footnote{See the footnote $(^3)$.}
\begin{equation} \frac{\partial R}{\partial
t_0}=-c,\qquad \frac{\partial x_0}{\partial t_0}=V\qquad{\rm and}\qquad
\frac{\partial V}{\partial t_0}=a.
\end{equation}

Now, using Eqs. (29) and (33), we want to verify the validity of
Eqs.(30)-(32).  The result of the verification is as follows\footnote{The
expressions (34)-(36) were calculated using the program ``Mathematica,
Version 2.2", therefore  it is easy to check these
calculations.}:

\begin{equation}
\frac{\partial E_z}{\partial t_0}\frac{\partial t_0}{\partial y}
-\frac{\partial
E_y}{\partial t_0}\frac{\partial t_0}{\partial z}=0,
\end{equation}

\begin{equation}
\frac{\partial E_x}{\partial t_0}\frac{\partial t_0}{\partial z}
-\frac{\partial
E_z}{\partial t_0}\frac{\partial t_0}{\partial x}+
\frac{1}{c}\frac{\partial B_y}{\partial t_0}\frac{\partial t_0}
{\partial t}
=-\frac{ac(z-z_0)}{[cR-V(x-x_0)]^3},
\end{equation}

\begin{equation}
\frac{\partial E_y}{\partial t_0}\frac{\partial t_0}{\partial x}
-\frac{\partial
E_x}{\partial t_0}\frac{\partial t_0}{\partial y}+
\frac{1}{c}\frac{\partial B_z}{\partial t_0}\frac{\partial t_0}
{\partial t}
=\frac{ac(y-y_0)}{[cR-V(x-x_0)]^3}.
\end{equation}

The
verification\footnote{There  is another manner to verify the validity of
Eqs. (30)-(32). If one substitutes {\bf E} and {\bf B} from (10) in Eq.
(7), one only has to satisfy oneself that the operators ``$\nabla\times$"
and ``$\partial/\partial t$" commute. In our case, because of ${\bf
V}=(V,0,0)$ and ${\bf A}=(A_x,0,0)$, it means the commutation of the
operators $\partial/\partial y({\rm or}\;z)$ and $\partial/\partial t$.
The verification shows that these operators do not commute if one uses
the rules (11).} shows that Eq.(30) is valid. But instead of
Eq.(31) and Eq.(32) we have Eq.(35) and Eq.(36) respectively. A reader has
to agree that this result is rather unexpected.

However, another way to obtain the fields (1) and (2) exists.  If one uses
{\it this} manner to verify the validity of Maxwell's equations, one finds
that fields (1) and (2) satisfy these equations.  In the next section we
shall consider this way in detail and we shall show that it does  not
correspond to a {\it pure} retarded interaction between the charge and the
point of observation.

$$$$
{\bf 4. Double (implicit and explicit) dependence   of $\varphi$, A, E
 and B on $t$ and $x_i$. Total derivatives: mathematical and physical
 aspects}

Let us, at the beginning, consider in detail Landau's method [2]  to obtain
the derivatives $\partial t_0/\partial t$ and $\partial t_0/\partial x_i$.
Landau considered two different expressions of the  function $R$:
\begin{equation}
R=c(t-t_0),\qquad {\rm where}\qquad t_0=f(x,y,z,t)
\end{equation}
and
\begin{equation}
R=[(x-x_0)^2+(y-y_0)^2+(z-z_0)^2]^{1/2},\qquad{\rm where}\qquad
x_{0i}=f_i(t_0).
\end{equation}
Then one calculates the derivatives ($\partial/\partial t$ and
$\partial/\partial x_i$) of functions (37) and (38), and
equating the results obtains  $\partial t_0/\partial t$ and $\partial
t_0/\partial x_i$. While  Landau uses here  a symbol $\partial$ (see
the expression before Eq. (63.6) in [2]) in order to emphasize that $R$
depends also on others independent variables $x,y,z$, it is easy to show
that he calculates here {\it total} derivatives of the functions (37),
(38) {\it with respect to} $t$ and $x_i$. The point is that if a {\it
given} function is expressed by two different types of functional
dependencies, exclusively {\it total} derivatives of these expressions
with respect to a {\it given} variable can be equated (contrary to the
{\it partial} ones). Here we adduce the scheme\footnote{In this scheme we have used
a symbol $d$ for a total derivative. In original text [2] we have
$$
\frac{\partial R}{\partial t}=
\frac{\partial R}{\partial t_0}\frac{\partial t_0}{\partial t}=
-\frac{{\bf RV}}{R}\frac{\partial t_0}{\partial t}=
c\left(1-\frac{\partial t_0}{\partial t}\right),
\qquad
{\rm and}
\qquad
\nabla t_0=-\frac{1}{c}\nabla R(t_0)=-\frac{1}{c}\left(
\frac{\partial R}{\partial t_0}\nabla t_0+\frac{{\bf R}}{R}\right).
$$} which was used in [2] to
 obtain $\partial t_0/\partial t$ and $\partial t_0/\partial x_i$:

\bigskip
\bigskip

\begin{equation}
\left[
\begin{array}{ccccc}
\underbrace{\frac{\partial R}{\partial t}_{(=c)}
+\frac{\partial R}{\partial t_0}_{(=-c)}
\frac{\partial t_0}{\partial t}}&=&
\underbrace{\frac{dR}{dt}}&=&
\underbrace{\sum\limits_k\frac{\partial R}{\partial x_{0k}}\frac{\partial
x_{0k}}{\partial t_0}_{\left(=-\frac{{\bf RV}}{R}\right)} \frac{\partial
t_0}{\partial t}}\\
\uparrow& &\uparrow& &\uparrow\\ R\{t,t_0(x_i,t)\}
&=&R(t_0)&=& R\{x_i,x_{0i}[t_0(x_i,t)]\}\\ \Updownarrow& &\Updownarrow&
&\Updownarrow\\
c(t-t_0)&=&R(t_0)&=&\left\{\sum_{i}[(x_i-x_{0i}(t_0)]^2\right\}^{1/2}\\
\downarrow& &\downarrow& &\downarrow\\
\overbrace{\frac{\partial R}{\partial t_0}_{(=-c)}
\frac{\partial t_0}{\partial x_i}}&=&
\overbrace{\frac{dR}{dx_i}}&=&
\overbrace{\frac{\partial R}{\partial
x_i}_{(=\frac{x_i-x_{0i}}{R})} + \sum\limits_k\frac{\partial R}{\partial
x_{0k}}\frac{\partial x_{0k}}{\partial t_0}_{\left(=-\frac{{\bf
RV}}{R}\right)} \frac{\partial t_0}{\partial x_i}} \end{array}\right]
\end{equation}

\bigskip
\bigskip

If one takes into account that $\partial t/\partial x_i=\partial
x_i/\partial t=0$, as a result  obtains the same values of the
derivatives which have been obtained in (14).

Let us now, as it was mentioned above in the fin of Section 3, calculate
the expressions (10) taking into consideration that the functions
$\varphi$ and {\bf A} depend on $t$ (or on $x_i$)\footnote{This depends on
the choice of the expression for $R$ in (37), (38)} implicitly and
explicitly {\it simultaneously}. In this case we have:  \begin{equation}
\frac{\partial\varphi}{\partial x_i}=-\frac{q}{(R-{\bf R}
\mbox{\boldmath$\beta$})^2}\left(\frac{\partial R}{\partial
x_i}-\frac{\partial{\bf R}}{\partial x_i}\mbox{\boldmath$\beta$}-{\bf
R}\frac{\partial\mbox{\boldmath$\beta$}}{\partial x_i}\right),
\end{equation}
\begin{equation}
\frac{\partial\varphi}{\partial t}=-\frac{q}{(R-{\bf R}
\mbox{\boldmath$\beta$})^2}\left(\frac{\partial R}{\partial
t}-\frac{\partial{\bf R}}{\partial t}\mbox{\boldmath$\beta$}-{\bf
R}\frac{\partial\mbox{\boldmath$\beta$}}{\partial t}\right),
\end{equation}
and
\begin{equation}
\frac{\partial {\bf A}}{\partial t}=\frac{\partial\varphi}{\partial t}
\mbox{\boldmath$\beta$}+
\varphi\frac{\partial\mbox{\boldmath$\beta$}}{\partial t},
\end{equation}
where
\begin{equation}
\frac{\partial\mbox{\boldmath$\beta$}}{\partial t}=
\frac{\partial\mbox{\boldmath$\beta$}}{\partial t_0}\frac{\partial
t_0}{\partial t}\qquad{\rm and}\qquad
\frac{\partial\mbox{\boldmath$\beta$}}{\partial x_i}=
\frac{\partial\mbox{\boldmath$\beta$}}{\partial t_0}\frac{\partial
t_0}{\partial x_i}.
\end{equation}

Now, let us consider {\it all}  derivatives in (10), (40)-(43)
 as {\it total} derivatives with respect to $t$
and $x_i$. Then, if substitute the expressions (40)-(43) in (10) (of
course, taking into account either {\it lhs} or {\it rhs} of the scheme
(39)), we obtain formally the {\it same} expressions for the fields (1)
and (2)!  Then if one substitutes the fields (1) and (2) in the Maxwell's
equation (7), considering {\it all} derivatives in (7) as {\it total}
ones and, of course, considering the functions {\bf E} and {\bf B} as
functions with both implicit and explicit dependence on $t$ (or on $x_i$),
one can see that the equation (7) is satisfied!

$$$$
{\bf 5. Conclusion}

If we consider {\it only} the implicit functional dependence of {\bf E}
and {\bf B} with respect to the time $t$, this means that we describe {\it
exclusively} the retarded interaction: the electromagnetic perturbation
created by the charge at the instant $t_0$ {\it reaches} the point of
observation $(x,y,z)$ after the time $\tau=R(t_0)/c$. Surprisingly, the
Maxwell equations {\it are not} satisfied in this case!

If we take into account a possible {\it explicit} functional dependence of
{\bf E} and {\bf B} with respect to the time $t$, {\it together} with the
{\it implicit} dependence, the Maxwell equations are satisfied. The
explicit dependence of {\bf E} and {\bf B} on $t$ means that, contrary to
the implicit dependence, {\it there is not} a retarded time for
electromagnetic perturbation {\it to reach} the point of observation. A
possible interpretation may be an action-at-a-distance phenomenon, as a
full-value solution of the Maxwell equations within the framework of the
so called ``dualism concept" [9,10].  In other words, there is a {\it
   simultaneous} and {\it independent} coexistence of  {\bf instantaneous
   long-range}  and  {\bf retarded short-range interactions}
   which cannot be reduced to each other.

$$$$
{\bf Acknowledgments}

We are grateful to Professor V. Dvoeglazov and Drs. D.W.Ahluwalia and
F.Brau for many stimulating discussions and critical comments.  We
acknowledge the paper of Professor E.Comay, which put an idea into us to
make the present work.

 \end{document}